\documentclass{ifacconf}

\usepackage{graphicx}
\usepackage{natbib}
\usepackage{amsmath,amssymb,amsfonts}
\usepackage{bm}
\usepackage{algorithm}
\usepackage{algpseudocode}
\usepackage{xcolor}
\usepackage{booktabs}
\usepackage{enumitem}
\usepackage{arydshln}
\usepackage{textcomp}
\usepackage{multirow}
\usepackage{accents}
\usepackage{caption}
\usepackage{subcaption}
\usepackage{balance}     
\usepackage{url}

\newtheorem{lemma}{Lemma}

\newtheorem{remark}{Remark}

\newcommand{\E}{\mathbb{E}}
\newcommand{\R}{\mathbb{R}}

\newcommand\norm[1]{\|#1\|}
\newlength{\dhatheight}

\newcommand{\U}{\mathcal{U}}
\newcommand{\IF}{I_{F,N}}

\newcommand{\hh}{\hat{h}}

\newcommand{\Var}{\mathbb{V}\mathrm{ar}}

\newcommand{\Tr}[1]{\text{Tr}\left\{#1\right\}}

\DeclareMathOperator*{\argmin}{arg\,min}

\begin{document}
\begin{frontmatter}

\title{Fundamental Limits of Data-Driven Predictive Control\thanksref{footnoteinfo}}

\thanks[footnoteinfo]{This work was supported by VINNOVA Competence Center AdBIOPRO 2016-05181, and  the Swedish Research Council through project 2025-04833.}

\author[KTH]{Jiabao He}
\author[ETH]{Feiran Zhao}
\author[KTH]{Yushan Li}
\author[KTH]{Yue Ju}
\author[ETH]{Florian D\"orfler}
\author[KTH]{H\r{a}kan Hjalmarsson}

\address[KTH]{Department of Decision and Control Systems, EECS, KTH Royal Institute of Technology,
100 44 Stockholm, Sweden (e-mail: jiabaoh, yushanl, yuej, hjalmars@kth.se).}
\address[ETH]{Automatic Control Laboratory, ETH Z\"urich, Physikstrasse 3, 8092 Z\"urich, Switzerland (e-mail: zhaofe@control.ee.ethz.ch, dorfler@ethz.ch).}

\begin{abstract}
		Data-driven predictive control (DDPC) has attracted substantial interest, and many variants have been proposed; however, comparatively less is known about the statistical closed-loop performance of these methods and the fundamental limits they are subject to. In this contribution, we apply the statistical decision framework developed in our earlier work \citep{He2026fundamental} to study fundamental limits of DDPC. The performance of a data-driven controller is measured by its risk, the expected deterioration in control performance relative to an oracle model-based controller at a given system, and by its total risk that measures the overall performance across all systems of interest. Lower bounds on these risks are derived using the bias–variance decomposition and the Cramér–Rao inequality. In particular, a lower bound on the total risk is obtained by a calculus-of-variations method. This bound applies to any DDPC design, thereby quantifying unavoidable performance limitations and providing a benchmark for assessing existing DDPC methods. In an illustrative example, we compare the risks of certainty-equivalence model predictive control, subspace predictive control, and data-enabled predictive control with the derived lower bounds, providing a quantitative performance assessment of these methods and some guidance on tuning regularization parameters in DDPC.
\end{abstract}


\end{frontmatter}
\vspace{-2mm}
\section{Introduction} \label{Sct1}

Data-driven control (DDC), the task of designing controllers directly from process data, has been studied for more than half a century \citep{Gevers93towards,Hjalmarsson2005experiment} and has recently witnessed a renewed surge of interest, driven by advances at the intersection of control, machine learning, optimization and related fields. In particular, two canonical benchmark problems have attracted considerable attention: the linear quadratic regulator (LQR) \citep{Fazel2018global,Recht2019tour,Dean2020sample,Mania2019certainty,De2019formulas,Van2020data,Dorfler2023certainty,Zhao2025regularization,Zhao2025data,Zeng2025noise}, and data-driven predictive control (DDPC) \citep{Coulson2019data,Berberich2020data,Dorfler2022bridging,Breschi2023data,Smith2024optimal,Chiuso2025harnessing,Sasfi2026gaussian}. We refer to \cite{Dorfler2023dataA,Dorfler2023dataB} for many more contributions. 

Most of these works focus on the design and analysis of particular controllers, and comparisons of performance of these controllers often lead to the answer ``it depends'' \citep{Dorfler2023dataB}. Since the advantage of one method over another is typically demonstrated by showing that its performance is closer to that of the optimal model-based controller, a more foundational question arises: is there a limit to how close \emph{any} data-driven controller can get to this oracle model-based controller? Answering this question is important, because such a limit points directly to the inescapable hardness of the problem, thereby clarifying what is achievable, and conversely what is not achievable, in DDC. In system identification, such a limit typically serves as a benchmark for assessing the efficiency of algorithms, such as the prediction error method \citep{Ljung1999system}, subspace identification \citep{He2024weighted} and identification with quantized observations \citep{Wang2024asymptotically}. In DDC, fundamental limits of this kind exist for the regret of online LQR \citep{Simchowitz2020naive,Ziemann2024regret,Bartos2026optimistic} and, in a local minimax sense, for offline LQR \citep{Lee2023fundamental} and LQG \citep{Lee2026fragility}. In our earlier work \cite{He2026fundamental}, we formulated explicit DDC as a statistical decision problem, treating a controller as a decision rule whose risk measures the expected performance degradation relative to the oracle model-based controller. We then derived risk lower bounds and illustrated the approach through case studies, including the benchmark problem of infinite-horizon LQR.

In this paper, we turn to another benchmark problem, finite-horizon DDPC. Whereas the infinite-horizon LQR has an explicit time-invariant optimal feedback policy, the optimal policy considered in DDPC is time-varying, depending on the initial condition, and often simply an implicit time series. These  features make DDPC policies harder to learn in general and introduce additional challenges in characterizing their fundamental limits. The main contributions of this paper are:
\begin{itemize}[leftmargin=*]
	\item We formulate the finite-horizon DDPC within the statistical decision framework of \cite{He2026fundamental} and derive a risk lower bound for any DDPC design. This bound reveals quantitative limitations that no controller design can avoid.
	\item We derive explicit gain sequences for some representative DDPC methods under a shrinking-horizon implementation and evaluate their performance against the derived lower bounds. This comparison quantifies how far these methods are from the fundamental limits. 
	\item In an illustrative example we show that there is no uniformly best DDPC over the prescribed domain. Moreover, the total risk lower bound implies a ``waterbed'' effect: any pointwise risk below the lower bound in one parameter region must be compensated by excess risk elsewhere, analogous to the fundamental tradeoff in classical sensitivity shaping.
\end{itemize}

The remainder of this paper is organized as follows. In Section~\ref{Sct2}, we introduce the statistical decision framework for DDC and derive lower bounds on risks. In Section~\ref{Sct3}, we introduce the DDPC problem, formulate it within this framework, and derive the explicit gain sequences of some representative DDPC designs. Section~\ref{Sct4} illustrates the lower bounds and the comparison of these designs on  first-order systems. Section~\ref{Sct5} concludes the paper.

\vspace{-2mm}
\section{Fundamental Limits of Data-Driven Control} \label{Sct2}
\vspace{-2mm}
\subsection{The Data} \label{Sct2.1}

Consider a dynamical system whose output $y_t$ is generated as
\begin{equation} \label{dynamical_system}
	y_t = g_t(Z^{t-1})+e_t, \quad y_0 = e_0,
\end{equation}
where $g_t(\cdot)$ is a deterministic function, $\{e_t\}$ is a sequence of independent random vectors with probability density functions (pdfs) $p_t(e)$ and has a zero mean, and $Z^{t-1}$ represents all past inputs $\{u_t\}$  and outputs $\{y_t\}$ up to the time $t-1$. The functions $g_t$ and pdfs $p_t$ belong to a parametrized family $\left\{\{g_t(\cdot;\theta),p_t(e;\theta)\}_{t=1}^{\infty}: \theta\in \mathcal{D}_\theta\right\}$, where $\theta \in \R^{n_\theta}$ is an unknown parameter, and $\mathcal{D}_\theta$ is an open set, representing the parameter domain of interest. 

The observed  data from a single trajectory of system \eqref{dynamical_system} that is available for a DDC problem is denoted by
\begin{equation}
	Z^{N}=\left\{u_0^d,y_0^d,\dots,u_{N-1}^d,y_{N-1}^d,y_{N}^d\right\},
\end{equation}
where $N$ denotes the number of observations. Using independence and the chain rule of probability, we have
\begin{equation} \label{eq:Bayes_pdfs}
	p(Z^N;\theta)=\prod_{t=1}^{N} p_t(\varepsilon_t(Z^t;\theta);\theta),
\end{equation}
where $\varepsilon_t(Z^t;\theta) =y_t-g_t(Z^{t-1};\theta)$ is the one-step prediction error. Then, the Fisher information matrix (FIM), quantifying the amount of information that the data $Z^N$ contains about the unknown $\theta$, is defined as
\begin{equation}\label{eq:fisher}
	\IF(\theta)=\E_\theta\!\Big[\frac{\partial\log p(Z^N;\theta)}{\partial\theta}\frac{\partial\log p(Z^N;\theta)}{\partial\theta^\top}\Big],
\end{equation}
where $\E_\theta[\cdot]$ denotes the expectation w.r.t. $p(Z^N;\theta)$. Throughout this work, we assume that the likelihood $\log p(Z^N;\theta)$ satisfies some standard regularity conditions, and $I_{F,N}(\theta)$ is positive definite for every $\theta\in\mathcal{D}_\theta$.

\subsection{A Statistical Decision Framework} \label{Sct2.2}

We now introduce a statistical decision framework for DDC.  For a given control task, let the oracle model-based control policy be $h(\theta): \mathbb{R}^{n_\theta} \to \mathbb{R}^{n_h}$, and a DDC policy be $\hat{h}  = \hat{h}(Z^N)$. We focus on a quadratic loss function,  and assume that $L_\theta\left(h(\theta)\right) = 0$, and write the loss as
\begin{equation} \label{loss}
	L_\theta(\hat{h}) = \big(\hat{h} - h(\theta)\big)^\top W(\theta) \big(\hat{h} - h(\theta)\big),
\end{equation}
where $W(\theta) \succ 0$ is a problem-dependent weighting matrix.

Since $\hat{h}$ is a function of $Z^N$, it is a random variable. The long-term performance of $\hat{h}$ is given by the risk
\begin{equation} \label{risk}
	R_\theta(\hat{h}) := \mathbb{E}_\theta\left[L_\theta(\hat{h})\right] = \int L_\theta(\hat{h})p(Z^N;\theta)dZ^N.
\end{equation}
Moreover, the performance of $\hat{h}$ over all possible parameters in $\mathcal{D}_\theta$ is captured by the total risk
\begin{equation} \label{average_risk}
	r(\hat h)=\int_{\mathcal{D}_\theta} R_\theta(\hat h) d\theta.
\end{equation}

The goal of DDC design is to choose a decision $\hat h(Z^N)$ with small risks. We refer to our earlier work \citep{He2026fundamental} and references therein for some design principles and methods. Instead of proposing a particular DDC design, we investigate how closely any DDC design can approach the oracle policy $h(\theta)$ using the available offline data $Z^N$. For this purpose, lower bounds on the pointwise risk $R_\theta(\hat h)$ and the total risk $r(\hat h)$ are derived in the next subsection.
\vspace{-2mm}
\subsection{Fundamental Limits}  \label{Sct2.3}

Define the bias and variance of a DDC policy $\hat h$ by
\begin{subequations} \label{risk-bias}
	\begin{align}
		b_\theta(\hat h)
		&:=\mathbb{E}_\theta[\hat h]-h(\theta),\\
		\Var_{\theta}(\hat{h})
		&:=\mathbb{E}_\theta\!\left[
		\bigl(\hat h-\mathbb{E}_\theta[\hat h]\bigr)
		\bigl(\hat h-\mathbb{E}_\theta[\hat h]\bigr)^\top
		\right].
	\end{align}
\end{subequations}
Using the bias--variance decomposition, we have
\begin{equation} \label{risk-decomposition}
	R_\theta(\hat h)
	=\|b_\theta(\hat h)\|_{W(\theta)}^2
	+\operatorname{Tr}\!\left\{
	W(\theta)\Var_{\theta}(\hat{h})\right\},
\end{equation}
where $\norm{b_\theta(\hat{h})}_{W(\theta)}^2 = {b_\theta(\hat{h})^\top W(\theta)b_\theta(\hat{h})}$, and $\Tr{\cdot}$ denotes the trace of a matrix. A standard approach to lower-bounding $R_\theta(\hat h)$ is to apply the Cram\'er--Rao inequality to $\Var_\theta(\hat h)$ for a prescribed bias function. To illustrate this, write $b_\theta:=b_\theta(\hat h)$ and $m_\theta:=\mathbb{E}_\theta[\hat h]$. Their Jacobians satisfy
\begin{equation*}
	m'_\theta =\frac{\partial m_\theta}{\partial\theta^\top}
	=b_\theta'+h'(\theta) \in\mathbb{R}^{n_h\times n_\theta}.
\end{equation*}
Combining \eqref{risk-decomposition} with the biased Cram\'er--Rao inequality yields \cite[Thm.~4.1]{He2026fundamental}
\begin{equation} \label{risk-lower-bound}
	R_\theta(\hat h) \geq \|b_\theta\|_{W(\theta)}^2
	+\operatorname{Tr}\!\left\{	W(\theta)m'_\theta I_{F,N}^{-1}(\theta)(m'_\theta)^\top \right\}.
\end{equation}
For a class of decision rules with the same bias $b_\theta$, \eqref{risk-lower-bound} provides a lower bound for the pointwise risk $R_\theta(\hat h)$. If $\hat h$ is unbiased, i.e., $b_\theta=b_\theta'=0$, \eqref{risk-lower-bound} then reduces to the unbiased Cram\'er--Rao lower bound (UCRLB):
\begin{equation} \label{risk-lower-bound_unbiased}
	R_\theta(\hat h)
	\geq\operatorname{Tr}\!\left\{
	W(\theta)h'(\theta)I_{F,N}^{-1}(\theta)h'(\theta)^\top
	\right\}.
\end{equation}
Ideally, to obtain the tightest possible risk bound, we would like to minimize the right-hand side in \eqref{risk-lower-bound} over all possible bias vectors. For every fixed value of $\theta$ at $\bar\theta$, taking $b_\theta=h(\bar\theta) - h(\theta)$ drives the right-hand side in \eqref{risk-lower-bound} to zero at $\theta=\bar\theta$. Clearly, this controller is not useful for values other than $\bar\theta$. Furthermore, $h(\theta)$ optimizes the bound for any specific $\theta = \bar\theta$, but not for all values of $\theta$. Thus, in general we cannot minimize the pointwise lower bound in \eqref{risk-lower-bound} uniformly for all $\theta\in \mathcal{D}_\theta$. 

A common benchmark for unrestricted-bias designs is instead obtained by integrating the bound \eqref{risk-lower-bound} over $\theta$ and optimizing over all admissible bias functions, which provides a lower bound on the total risk $r(\hat h)$ in \eqref{average_risk}. To illustrate this, denote the right-hand side in \eqref{risk-lower-bound}  as
\begin{equation*}
	f(\theta,b_\theta,b_\theta') :=\norm{b_\theta}_{W(\theta)}^2 + \Tr{W(\theta)m'_\theta I_{F,N}^{-1}(\theta)(m'_\theta)^\top} .
\end{equation*}
Then, we formulate a functional minimization problem
\begin{equation} \label{Bayes-risk-CoV-scalar}
	\min_{b_\theta} \quad F(b_\theta) := \int_{\mathcal{D}_\theta} f(\theta,b_\theta,b_\theta') d\theta.
\end{equation}
By \eqref{average_risk}, the solution to \eqref{Bayes-risk-CoV-scalar} provides a lower bound on the total risk $r(\hat h)$ of any DDC $\hat h$. It also determines the optimal bias $b_\theta^*$ that attains this bound. To characterize the solution of the functional minimization problem \eqref{Bayes-risk-CoV-scalar}, we assume that the admissible bias $b_\theta$ is twice continuously differentiable. Since $W(\theta)$ and $I_{F,N}(\theta)$ are positive definite, the Lagrangian $f$ is strictly convex in $(b_\theta,b_\theta')$. Therefore, if the minimizer exists, it is unique, and can be characterized by the Euler--Lagrange equation and the natural boundary condition; see Thm.~4.2 in \cite{He2026fundamental}. These conditions generally form a set of boundary-value partial differential equations in $b_\theta$, and can be solved numerically. Further details on computing the lower bound are provided in Section~\ref{Sct4}.

\vspace{-2mm}
\section{Data-Driven Predictive Control} \label{Sct3}
\vspace{-2mm}
\subsection{Problem Setup} \label{Sct3.1}

We now introduce a DDPC problem and cast it into the statistical decision framework of Section~\ref{Sct2}. For simplicity, we focus on unconstrained linear quadratic control, which serves as a prototypical model predictive control problem. Extensions to generalized predictive control \citep{Bitmead1990adaptive}, such as constrained problems and partially observed systems are left for future work. Consider the following stochastic linear quadratic control problem over a finite horizon $T$:
\begin{align}
	\min_{\left\{u_k\right\}_{k=0}^{T-1}} \quad J_T &= \E\Big[\sum_{k=0}^{T-1}\left(x_k^\top Q x_k+u_k^\top R u_k\right) + x_T^\top Q_T x_T\Big], \label{eq:MPC}\\
	\mathrm{s.t.}\quad
	x_{k+1} &=Ax_k+Bu_k+e_k, \label{eq:dynamics}
\end{align}
where $x_k\in\mathbb{R}^{n_x}$ and $u_k\in\mathbb{R}^{n_u}$. The initial state is $x_0\sim\mathcal{N}(0,\Sigma_0)$, with $\Sigma_0\succ0$, and the process noises $e_k\sim\mathcal{N}(0,\sigma_e^2I_{n_x})$ are independent and identically distributed (i.i.d.), independent of $x_0$. The expectation in cost \eqref{eq:MPC} is over $x_0$ and process noise. Moreover, matrices $Q$ and $R$ are positive definite, and $Q_T$ is positive semidefinite.

When matrices $A$ and $B$ are known, the optimal control policy for \eqref{eq:MPC} admits an analytical form, which is summarized in the following lemma:

\begin{lemma}[Oracle policy {\citep{Bertsekas1995dynamic}}]
	\label{lemma:dp}
	Set \\ $P_T=Q_T$ and define, backward for $k=T-1,\ldots,0$,
	\begin{subequations} \label{eq:riccati}
		\begin{align}
			S_k&=R+B^\top P_{k+1}B, \label{eq:riccati_S}\\
			K_k&=-S_k^{-1}B^\top P_{k+1}A, \label{eq:riccati_K}\\
			P_k&=Q+A^\top P_{k+1}A
			-A^\top P_{k+1}BS_k^{-1}B^\top P_{k+1}A. \label{eq:riccati_P}
		\end{align}
	\end{subequations}
	The optimal policy for \eqref{eq:MPC} is
	\begin{equation} \label{eq:Kt}
		u_k=K_kx_k,\quad k=0,\ldots,T-1.
	\end{equation}
\end{lemma}
The gain sequence $\bm K:=(K_0,\ldots,K_{T-1})$ represents the oracle model-based control. The corresponding cost \eqref{eq:MPC} under the policy \eqref{eq:Kt} is denoted by $J_T(\bm K)$.
\vspace{-2mm}
\subsection{DDPC in the Statistical Decision Framework}
\label{Sct3.2}

Suppose now that $A$ and $B$ are unknown and that the controller is designed from an offline trajectory of \eqref{eq:dynamics}:
\begin{equation*}
	Z^N =\{x_0^d,u_0^d,\ldots,x_{N-1}^d,u_{N-1}^d,x_N^d\},
\end{equation*}
where the excitation inputs are i.i.d., distributed as $u_k^d\sim\mathcal{N}(0,\sigma_u^2I_{n_u})$, and are independent of initial state and process noises. 

Since the optimal policy is in the form \eqref{eq:Kt}, we only consider DDPC designs that return a time-varying linear feedback policy,
\begin{equation} \label{eq:Kt_data_driven}
	u_k=\widehat K_kx_k,\quad k=0,\ldots,T-1,
\end{equation}
where $\widehat{\bm K}:=(\widehat K_0,\ldots,\widehat K_{T-1})$ is a function of $Z^N$. Correspondingly, the cost \eqref{eq:MPC} under the policy \eqref{eq:Kt_data_driven} for the true plant is denoted by $J_T(\widehat{\bm K})$. The next lemma relates the closed-loop performance gap $J_T(\widehat{\bm K})-J_T(\bm K)$ to the gain mismatch.
\begin{lemma}[Performance gap \citep{Hambly2021policy}] \label{lem:mismatch}
	Let \\ $x_k$ and $\hat x_k$ denote the states under gain sequences $\bm K$ and $\widehat{\bm K}$, respectively, with $x_0 = \hat x_0$. Define
	\begin{equation*}
		P_k = \Sigma_k:=\mathbb{E}[x_kx_k^\top],\
		\widehat\Sigma_k:=\mathbb{E}[\hat x_k\hat x_k^\top],\
		\widetilde K_k:=\widehat K_k-K_k.
	\end{equation*}
	Then, the gap satisfies
	\begin{equation} \label{eq:cost_gap}
		\begin{split}
			&J_T(\widehat{\bm K})-J_T(\bm K) 
			=\sum_{k=0}^{T-1}\operatorname{Tr}\!\left\{
			\widehat\Sigma_k \widetilde K_k^\top S_k\widetilde K_k
			\right\} \approx \\
			&\sum_{k=0}^{T-1}\operatorname{Tr}\!\left\{
			\Sigma_k\widetilde K_k^\top S_k\widetilde K_k
			\right\} + \mathcal{O}_p\left(\norm{\widehat{\bm K}-\bm K}^3\right),
		\end{split}
	\end{equation}
	provided that $\widehat {\bm{K}}$ is sufficiently close to ${\bm{K}}$ in the sense that $\widehat\Sigma_k \approx \Sigma_k + \mathcal{O}_p\left(\norm{\widehat {\bm{K}}-{\bm{K}}}\right)$, where $\norm{\cdot}$ denotes the spectral norm of a matrix, and $M_k =  \mathcal{O}_p(m_k)$ means that $M_k/m_k$ is stochastically bounded.
\end{lemma}

Let $\theta$ collect the unknown parameters in matrices $A$ and $B$. The lemma above motivates the following quadratic loss as a surrogate for the closed-loop performance gap:
\begin{equation} \label{eq:loss}
	\begin{split}
		L_\theta(\hat h)
		=&\sum_{k=0}^{T-1}\operatorname{Tr}\!\left\{
		\Sigma_k\widetilde K_k^\top S_k\widetilde K_k
		\right\}\\
		=&\bigl(\hat h-h(\theta)\bigr)^\top
		W(\theta)\bigl(\hat h-h(\theta)\bigr),
	\end{split}
\end{equation}
where $h(\theta)=\mathrm{Vec}(\bm K) \in \mathbb{R}^{n_h}$, $\hat h=\mathrm{Vec}(\widehat{\bm K}) \in \mathbb{R}^{n_h}$,  with $n_h=Tn_xn_u$. Here, $\mathrm{Vec}(\cdot)$ stacks the column-wise vectorizations of the gain matrices. The weighting matrix has the block-diagonal form
\begin{equation} \label{eq:W}
	W(\theta)
	:=\mathrm{blkdiag}\!\left(
	\Sigma_0\otimes S_0,\ldots,
	\Sigma_{T-1}\otimes S_{T-1}
	\right) \succ0,
\end{equation}
where $\otimes$ denotes the Kronecker product.
Under the loss \eqref{eq:loss}, the pointwise risk $R_\theta(\hat h)$ and total risk $r(\hat h)$ can be further defined according to \eqref{risk} and \eqref{average_risk}, respectively. Then, the corresponding lower bounds can be derived.

We next introduce two families of DDPC methods: namely, indirect and direct methods, depending on whether system matrices $A$ and $B$ are explicitly identified. For each method considered, we derive an explicit gain sequence $\widehat{\bm K}$.
\vspace{-2mm}
\subsection{Design of DDPC: Certainty-Equivalence Method} \label{Sct3.3}

The certainty-equivalence indirect method first estimates matrices $A$ and $B$ from $Z^N$ and then applies the model-based design in Lemma~\ref{lemma:dp} to the estimates. For estimation, the ordinary least-squares (OLS) estimate is often used:
\begin{equation}\label{eq:ls}
	\big[\widehat A\ \ \widehat B\big]
	=\argmin_{A,B}\sum_{k=0}^{N-1}\big\|x_{k+1}^d-Ax_k^d-Bu_k^d\big\|^2.
\end{equation}
With the certainty-equivalence  principle (CEP), the control law is then obtained by iterating the Riccati recursion in \eqref{eq:riccati} backwards with $\widehat A$ and $\widehat B$ in place of $A$ and $B$. Let $\widehat{\bm K}^{\mathrm{CE}}$ denote the resulting gain sequence in \eqref{eq:riccati_K} and $\hh^{\mathrm{CE}}$ its vectorization. The statistical properties of this method were studied by \cite{Ye2024online,Liu2026certainty}. It should be mentioned that this design treats the estimated model as exact, whereas many regularized DDPC methods seek to improve robustness to uncertainty arising from noisy data \citep{Dorfler2022bridging,Chiuso2025harnessing}.
\vspace{-2mm}
\subsection{Design of DDPC: Direct Methods} \label{Sct3.4}

Direct methods approximate the cost in \eqref{eq:MPC} without explicitly estimating system parameters $A$ and $B$. At each step, they minimize this approximation and apply the first optimized input. We use a shrinking horizon implementation, with the prediction horizon set to the remaining task horizon, and derive gain sequences explicitly. As emphasized in \cite{Anand2024data}, closed-loop optimality of DDPC is tied to the sequential decision structure rather than the indirect or direct distinctions. Therefore, these explicit gain sequences provide a new perspective for understanding different DDPC methods. They also allow these methods to be cast within our statistical decision framework, enabling comparisons with the oracle controller and the derived lower bounds.
\vspace{-2mm}
\subsubsection{\textbf{Multi-Step Predictor Model.}}

At time $k\le T-1$ the controller has measured $x_k$ and has $f_k:=T-k$ penalized outputs ahead, $x_{k+1},\dots,x_T$. First, we collect the future inputs and states into
\begin{equation*}\label{eq:stacked}
	\bm u_{f_k}:=\begin{bmatrix}u_k\\ \vdots\\ u_{T-1}\end{bmatrix}\in\R^{n_uf_k},
	\bm x_{f_k}:=\begin{bmatrix}x_{k+1}\\ \vdots\\ x_{T}\end{bmatrix}\in\R^{n_xf_k}.
\end{equation*}
Then, given $x_k$, the term $x_k^\top Qx_k$ in \eqref{eq:MPC} is fixed, and the cost-to-go from time $k$ reduces to
\begin{equation}\label{eq:ctg}
	J_k(\bm u_{f_k}):=\E\Big[ \norm{\bm x_{f_k}}_{Q_{f_k}}^2+\norm{\bm u_{f_k}}_{R_{f_k}}^2\Big|\ x_k\Big],
\end{equation}
where $Q_{f_k} =\mathrm{blkdiag}(Q,\dots,Q,Q_T)$, and $R_{f_k}=I_{f_k}\otimes R$. Meanwhile, stacking \eqref{eq:dynamics} over the horizon gives
\begin{equation}\label{eq:true-pred}
	\bm x_{f_k}=\Gamma_{f_k} x_k+G_{f_k} \bm u_{f_k} + H_{f_k} \bm e_{f_k},
\end{equation}
where
\begin{equation*}\label{eq:GHG}
	\Gamma_{f_k}=\begin{bmatrix}A\\ A^2\\ \vdots\\ A^{f_k}\end{bmatrix},
	G_{f_k}=\begin{bmatrix}B&0&\cdots&0\\ AB&B&\cdots&0\\ \vdots&\vdots&\ddots&\vdots\\ A^{f_k-1}B&A^{f_k-2}B&\cdots&B\end{bmatrix},
\end{equation*}
and $\bm e_{f_t}$ and $H_{f_k}$ are similarly defined as $\bm u_{f_k}$ and $G_{f_k}$, respectively. This multi-step relation is the starting point for many direct DDPC methods \citep{Favoreel1999spc,Breschi2023data,Chiuso2025harnessing}. Since $\Gamma_{f_k}$ and $G_{f_k}$ depend on the unknown system parameters, these methods construct a predictor that is linear in the initial state $x_k$ and future inputs $\bm u_{f_k}$:
\begin{equation} \label{eq:predictor}
	\hat{\bm x}_{f_k}=\widehat \Gamma_{f_k} x_k + \widehat G_{f_k} \bm u_{f_k},
\end{equation}
where $\widehat \Gamma_{f_k}$ and $\widehat H_{f_k}$ are
estimated from $Z^N$ without estimating matrices $A$ and $B$ explicitly. Then, they minimize the approximate cost-to-go
\begin{equation}\label{eq:ctg-hat}
	\widehat J_k(\bm u_{f_k}):=\norm{\widehat\Gamma_{f_k}x_k+\widehat G_{f_k}\bm u_{f_k}}_{Q_{f_k}}^2 + \norm{\bm u_{f_k}}_{R_{f_k}}^2 ,
\end{equation}
and apply the first element of the minimizer. To be specific, since \eqref{eq:ctg-hat} is a quadratic in $\bm u_{f_k}$, the minimizer  is
\begin{equation} \label{eq:DD-linear-law}
	\bm u_{f_k}^\star=-\big(\widehat G_{f_k}^\top Q_{f_k}\widehat G_{f_k}+R_{f_k}\big)^{-1}\widehat G_{f_k}^\top Q_{f_k}\widehat\Gamma_{f_k}x_k,
\end{equation}
so the applied first input is $u_k=\widehat K_kx_k$ with
\begin{equation}\label{eq:linear-law}
	\widehat K_k=-{V}_{f_k} \big(\widehat G_{f_k}^\top Q_{f_k}\widehat G_{f_k}+R_{f_k}\big)^{-1}\widehat G_{f_k}^\top Q_{f_k}\widehat\Gamma_{f_k}, 
\end{equation}
where ${V}_{f_k} :=\begin{bmatrix}I_{n_u}&0\end{bmatrix} \in \mathbb{R}^{n_u \times n_uf_k}$,  for $k=0,\dots,T-1$. If $\Gamma_{f_k}$ and $G_{f_k}$ are at their true values, the solution \eqref{eq:DD-linear-law} is identical to the oracle solution \eqref{eq:Kt}. 

\begin{remark}
    A key difference among existing direct DDPC methods is how they construct the multi-step predictor~\eqref{eq:predictor}, including exploiting structural properties to enforce causality \citep{Sader2025causality},  accounting for prediction uncertainty \citep{Smith2024optimal,Chiuso2025harnessing}, and eliminating bias when data $Z^N$ are gathered in closed-loop \citep{Moffat2024transient,Liu2025closed,Dinkla2026closed}.
\end{remark}

We next introduce two representative direct methods for comparison: Subspace Predictive Control (SPC) \citep{Favoreel1999spc} and Data-enabled Predictive Control (DeePC) \citep{Coulson2019data}. Both were originally formulated using input--output data. Since the state is available in our setting, the implementation details differ from those of the original methods, but the main ideas remain unchanged. Meanwhile, their original input--output formulations can be recovered by replacing the measured state with a suitable window of past inputs and outputs, viewed as a non-minimal state \citep{Premer2026trajectory}.

\vspace{-2mm}
\subsubsection{\textbf{SPC} \citep{Favoreel1999spc}:}

Define $N_T:=N-T+1$. For every horizon $f_k \le T$, arranging data in $Z^N$ column-wise gives the data Hankel matrices
\begin{equation}\label{eq:hankel}
	\begin{split}
		U_{f_k}^d&=\begin{bmatrix}u_0^d&u_1^d&\cdots&u_{N_T-1}^d\\ \vdots&\vdots&\ddots&\vdots\\ u_{f_k-1}^d&u_{f_k}^d&\cdots&u_{N_T+f_k-2}^d\end{bmatrix}\in\R^{n_uf_k\times N_T},\\
		X_{f_k}^d&=\begin{bmatrix}x_1^d&x_2^d&\cdots&x_{N_T}^d\\ \vdots&\vdots&\ddots&\vdots\\ x_{f_k}^d&x_{f_k+1}^d&\cdots&x_{N_T+f_k-1}^d\end{bmatrix}\in\R^{n_xf_k\times N_T},\\
		X_0^d &=\begin{bmatrix}x_0^d&x_1^d&\cdots&x_{N_T-1}^d\end{bmatrix}\in\R^{n_x\times N_T}.\\
	\end{split}
\end{equation}
By \eqref{eq:true-pred}, we have
\begin{equation}
	\begin{split}
		X_{f_k}^d &= \Gamma_{f_k}X_0^d +  G_{f_k} U_{f_k}^d + H_{f_k}E_{f_k}^d \\
		& = \begin{bmatrix} \Gamma_{f_k} & G_{f_k} \end{bmatrix} \Phi_{f_k}^d+H_{f_k} E_{f_k}^d,
	\end{split}
\end{equation}
where $\Phi_{f_k}^d = \begin{bmatrix}
	X_0^d \\U_{f_k}^d
\end{bmatrix}$, and $E_{f_k}^d$ is the corresponding noise matrix. SPC estimates the multi-step predictor \eqref{eq:predictor} directly by OLS
\begin{equation}\label{eq:spc-ls}
	\begin{bmatrix} \widehat\Gamma_{f_k} & \widehat G_{f_k} \end{bmatrix} =X_{f_k}^d(\Phi_{f_k}^d)^\top\big(\Phi_{f_k}^d(\Phi_{f_k}^d)^\top\big)^{-1},
\end{equation}
provided that $\Phi_{f_k}^d$ is full-row rank. Then, the gains are obtained using \eqref{eq:linear-law}, denoted by $\widehat K_k^{\mathrm{SPC}}$ and $\hh^{\mathrm{SPC}}$ their vectorization.
\vspace{-2mm}
\subsubsection{\textbf{DeePC} \citep{Coulson2019data}:}

DeePC  also uses the data matrices \eqref{eq:hankel}. By the fundamental lemma \citep{Willems2005note}, if the data was noise free and the input was persistently exciting of sufficient order, then $(x_k,\bm u_{f_k},\bm x_{f_k})$ would be a trajectory of \eqref{eq:dynamics}, if and only if
\begin{equation}\label{eq:fl}
	\begin{bmatrix}X_{0}^d\\ U_{f_k}^d\\ X_{f_k}^d\end{bmatrix}g=\begin{bmatrix}x_{k}\\ \bm u_{f_k}\\ \bm x_{f_k}\end{bmatrix}, \quad \text{for some }g\in\R^{N_T}.
\end{equation}
With noisy data, \eqref{eq:fl} is kept as a constraint and the vector $g$ is regularized \citep{Coulson2019data}. There are various choices for the regularization term, such as the $\ell_1$-norm, the squared $\ell_2$-norm, and the projection-based regularizer \citep{Dorfler2022bridging}. Here we use the projection-based regularizer, which penalizes only the component of $g$ outside the row space of $[X_0^d;\,U_{f_k}^d] =: \Phi_{f_k}^d$. Let $\Pi_{f_k}:=(\Phi_{f_k}^d)^\dagger\Phi_{f_k}^d$ be the orthogonal projector onto its row space, so that $I-\Pi_{f_k}$ projects onto the kernel of $\Phi_{f_k}^d$.  For every horizon $f_k \le T$, DeePC is then formulated as
\begin{equation}\label{eq:deepc}
	\begin{split}
		\min_{g,\bm u_{f_k}, \hat{\bm x}_{f_k}} & \norm{\hat{\bm x}_{f_k}}^2_{Q_{f_k}}+\norm{\bm u_{f_k}}^2_{R_{f_k}}+\lambda \norm{(I-\Pi_{f_k})g}^2, \\
		\text{s.t.}\quad &\begin{bmatrix}X_{0}^d\\ U_{f_k}^d\\ X_{f_k}^d\end{bmatrix}g=\begin{bmatrix}x_k\\ \bm u_{f_k}\\ \hat{\bm x}_{f_k}\end{bmatrix},
	\end{split}
\end{equation}
where $\lambda>0$. Compared with SPC, DeePC skips the explicit predictor estimation step, and it requires the predicted trajectory to be a linear combination of the recorded segments. The equivalence between SPC and DeePC in some special cases is discussed in \cite{Fiedler2021relationship,Mattsson2024equivalence}. In particular, as $\lambda\to\infty$ the regularizer confines $g$ to the row space of $\Phi_{f_k}^d$, and \eqref{eq:deepc} reduces to SPC \citep{Dorfler2022bridging}.

We now study the explicit solution of DeePC. Define
\begin{equation}
	\mathcal Q_{f_k}:=(X_{f_k}^d)^\top Q_{f_k}X_{f_k}^d+(U_{f_k}^d)^\top R_{f_k}U_{f_k}^d+\lambda (I-\Pi_{f_k}).
\end{equation}
Substituting $\bm u_{f_k}=U^d_{f_k}g$ and $\hat{\bm x}_{f_k}=X^d_{f_k}g$ turns \eqref{eq:deepc} into an equality-constrained quadratic program in $g$:
\begin{equation}\label{eq:deepc-g}
	\begin{split}
		\min_{g} \quad & g^\top\mathcal Q_{f_k}g,\\
		\text{s.t.} \quad &X_{0}^dg=x_k.
	\end{split}
\end{equation}
With a multiplier $\nu$, the Lagrangian is given by
\begin{equation}
	\mathcal L(g,\nu)
	=g^\top\mathcal Q_{f_k}g
	+\nu^\top(X_0^dg-x_k).
\end{equation}
Stationarity gives $g=-\frac12\mathcal Q_{f_k}^{-1}(X_0^d)^\top\nu$.
Substituting this expression into $X_0^dg=x_k$ yields
\begin{equation}
	\nu^\star
	=-2\Bigl(X_0^d\mathcal Q_{f_k}^{-1}(X_0^d)^\top\Bigr)^{-1}x_k.
\end{equation}
Substituting $\nu^\star$ back gives
\begin{equation}\label{eq:deepc-gstar}
	g^\star
	=\mathcal Q_{f_k}^{-1}(X_0^d)^\top
	\Bigl(X_0^d\mathcal Q_{f_k}^{-1}(X_0^d)^\top\Bigr)^{-1}x_k,
\end{equation}
so that the applied first input $u_k=\bar{\bm I} U_{f_k}^dg^\star$ is again a linear feedback $u_k=\hat K_k^{\mathrm{DeePC}}x_k$ with
\begin{equation}\label{eq:K-deepc}
	\hat K_k^{\mathrm{DeePC}}={V}_{f_k}U_{f_k}^d\,\mathcal Q_{f_k}^{-1}(X_{0}^d)^\top\Big(X_{0}^d\,\mathcal Q_{f_k}^{-1}(X_{0}^d)^\top\Big)^{-1}.
\end{equation}
We use $\hh^{\mathrm{DeePC}}$ to denote their vectorization.

\vspace{-2mm}
\subsection{Design of DDPC: Total Risk Minimization} \label{Sct3.5}

Finally, we introduce another method that minimizes the total risk
$r(\hat h)$ in \eqref{average_risk}. This is often referred as Bayes control in the literature \citep{Scampicchio2019bayesian,Ferizbegovic2021bayes,Chiuso2025harnessing}. Interchanging the order of integration gives
\begin{equation}
	r(\hat h)=\int\left(\int_{\mathcal{D}_\theta} L_\theta(\hat h)p(Z^N;\theta)d\theta\right)dZ^N.
\end{equation}
Hence, for each fixed observation $Z^N$, we are led to minimize
\begin{equation}
	V(\hat h) = \int_{\mathcal{D}_\theta} L_\theta(\hat h) p(Z^N;\theta)d\theta.
\end{equation}
For the quadratic loss $L_\theta(\hat h)$, the unconstrained minimizer of $V(\hat h)$ takes the form
\begin{equation}\label{eq:Bayes_control}
	\hat h^{\mathrm{A(P)}} :=\arg\min_{\hat h} V(\hat h) =\overline F^{-1}(Z^N) \overline D(Z^N),
\end{equation}
where the superscript $A(P)$ stands for the total (average) risk minimization, having the origin in the Pitman estimator \citep{Pitman1939estimation}, and $\overline F(Z^N):=\int_{\mathcal{D}_\theta}W(\theta)p(Z^N;\theta)\,d\theta$, and $\overline D(Z^N) :=\int_{\mathcal{D}_\theta}W(\theta)h(\theta)p(Z^N;\theta)\,d\theta$. Since $\hat h^{\mathrm{A(P)}}$ minimizes the integrand pointwise in $Z^N$, the resulting rule $\hat h^{\mathrm{A(P)}}$ minimizes the total risk $r(\hat h)$.  It should be mentioned that to compute $\hat h^{\mathrm{A(P)}}$, the integrals $\overline F(Z^N)$ and $\overline D(Z^N)$ have to be computed, which in general requires sampling methods.

\vspace{-2mm}
\section{Simulation}  \label{Sct4}

For ease of illustration, we now use scalar systems to demonstrate the effectiveness of our approach and benchmark the performance of those DDPC methods introduced in Section~\ref{Sct3}. The code is publicly available\footnote{ \url{https://github.com/jiabaoh/Fundamental-Limits-DDPC}.}.
\vspace{-2mm}
\subsection{Experiment Setup}\label{Sct4.1}

Given a control horizon $T=4$, consider the following linear quadratic control problem
\begin{subequations}\label{eq:JN}
	\begin{align}
		\min_{\{u_k\}_{k=0}^{T-1}}\quad J_T &= \E\Big[\sum_{k=0}^{T-1}\big(qx_k^2+ru_k^2\big)\Big], \label{eq:JN_MPC} \\
		\text{s.t.}\quad x_{k+1} &=ax_k+bu_k+e_{k+1}. \label{eq:JN_dynamics}
	\end{align}
\end{subequations}
This is the finite-horizon DDPC problem introduced in Section~\ref{Sct3} with $A=a$, $B=b$, $Q=q$, $R=r$, $\Sigma_0=\sigma_0^2$, and the absence of a terminal term corresponding to $Q_T=0$. As in \cite{He2026fundamental}, we assume that all systems of interest lie in a domain
\begin{equation*}
	\mathcal{D}_\theta = \left\{\theta : (a-a_\circ)^2 + (b- b_\circ)^2 \le \rho_\circ^2 \right\},
\end{equation*}
where $a_\circ = b_\circ = 0.5$ and $\rho_\circ = 0.4$. Moreover, we take $q = 1$ and $r = 0.1$. For the offline data collection, we take the sample size $N = 20$, and $\sigma_e^2 = \sigma_0^2 = \sigma_u^2 = 1$. For each grid point $\theta_i = (a_i, b_i)$ in $\mathcal{D}_\theta$, we perform 1000 Monte Carlo trials and use their average to estimate the risk and suboptimality gap of each realized DDPC, inlcuding CEP, SPC, DeePC and the total risk minimization method, denoted by $\hat{h}^\mathrm{CE}$, $\hat{h}^\mathrm{SPC}$, $\hat{h}^\mathrm{DeePC}$ and $\hat{h}^\mathrm{A(P)}$, respectively.

\subsubsection{\textbf{Computing the bounds}:} Two fundamental lower bounds are computed for illustration. The first is the UCRLB \eqref{risk-lower-bound_unbiased} for the pointwise risk $R_\theta(\hat h)$ of unbiased DDPC, and the second is the optimal-bias lower bound on the total risk $r(\hat h)$ that applies to all controllers. Computation of these two bounds require the sensitivities of the oracle gains $\bm{K}$ and the Fisher information matrix. The former are obtained by differentiating the Riccati recursion of Lemma~\ref{lemma:dp} analytically, and the latter is computed from the recursion of the state variance. We refer to the Appendix  in \cite{He2026fundamental} for details of similar derivations. Since both $I_{F,N}(\theta)$ and $W(\theta)$ are block diagonal, the integrand $f(\theta,b_\theta,b_\theta') $ of the optimal-bias bound is a sum over $t$ of terms that involve only the bias of the $t$-th gain, so the variational problem \eqref{Bayes-risk-CoV-scalar} decouples into $T$ independent problems, and $F(b_\theta^*)$ is the sum of their minima. Each of them is solved through its Euler--Lagrange equation, a linear second-order partial differential equation. Moreover, since the bias is not constrained on the boundary of $\mathcal D_\theta$, the minimization is subject to natural boundary conditions. We solve the equation numerically with the finite element method using the Partial Differential Equation Toolbox of MATLAB, where the parameter domain $\mathcal D_\theta$ is covered by a triangular mesh. The solver returns $b_\theta^*$ and its gradient at the mesh nodes, from which $f(\theta,b_\theta^*,(b_\theta^*)^{\prime})$ is evaluated and $F(b_\theta^*)$ is obtained by numerical integration over the mesh, and the UCRLB is evaluated at the same nodes. 

\vspace{-2mm}
\subsection{Simulation Results} \label{Sct4.2}

\begin{figure}
	\centering
	\includegraphics[scale=0.14]{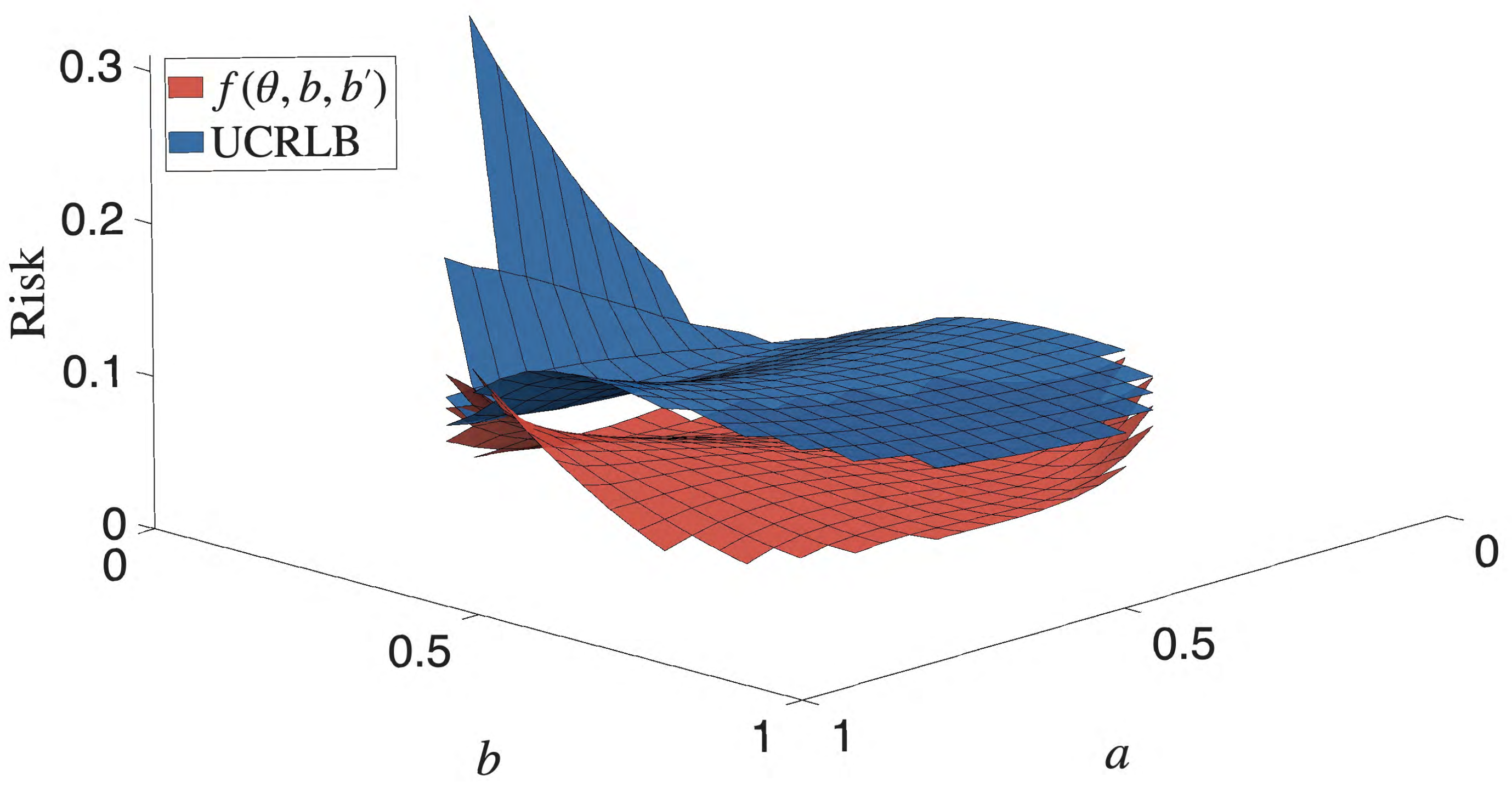}
	\caption{Two lower bounds for DDPC over $\mathcal{D}_\theta$.}
	\label{fig:ex3-two-bounds}
\end{figure}

Figure~\ref{fig:ex3-two-bounds} shows the UCRLB \eqref{risk-lower-bound_unbiased} (blue) and the integrand $f(\theta,b_\theta^*,b_\theta^{*\prime})$ evaluated at the optimal bias (red) over $\mathcal{D}_\theta$. The UCRLB bounds the pointwise risk of unbiased DDPC controllers, whereas the integral of the red surface bounds the total risk of any DDPC controller, including biased designs. Since zero bias is admissible in the variational problem, the optimal-bias bound cannot exceed the integrated UCRLB. As in the infinite-horizon case \citep{He2026fundamental}, both bounds rise sharply when $b$ is small and $a$ is close to $1$. The reason is that for $|a| < 1$, the Fisher information grows linearly with $N$ and is approximately given by $I_{F,N}(\theta)\approx\frac{N}{\sigma_e^2}\,{\rm diag}\big(\frac{b^2\sigma_u^2+\sigma_e^2}{1-a^2},\,\sigma_u^2\big)$ for large $N$, which decreases as $b\to0$ and $|a|\to1$. Since the bounds depend on the inverse of the Fisher information matrix, they increase accordingly. This verifies our earlier observation in \cite{He2026fundamental}: systems that are hard to identify are likewise hard to learn to control from data.

\begin{figure}
	\centering
	\includegraphics[scale=0.12]{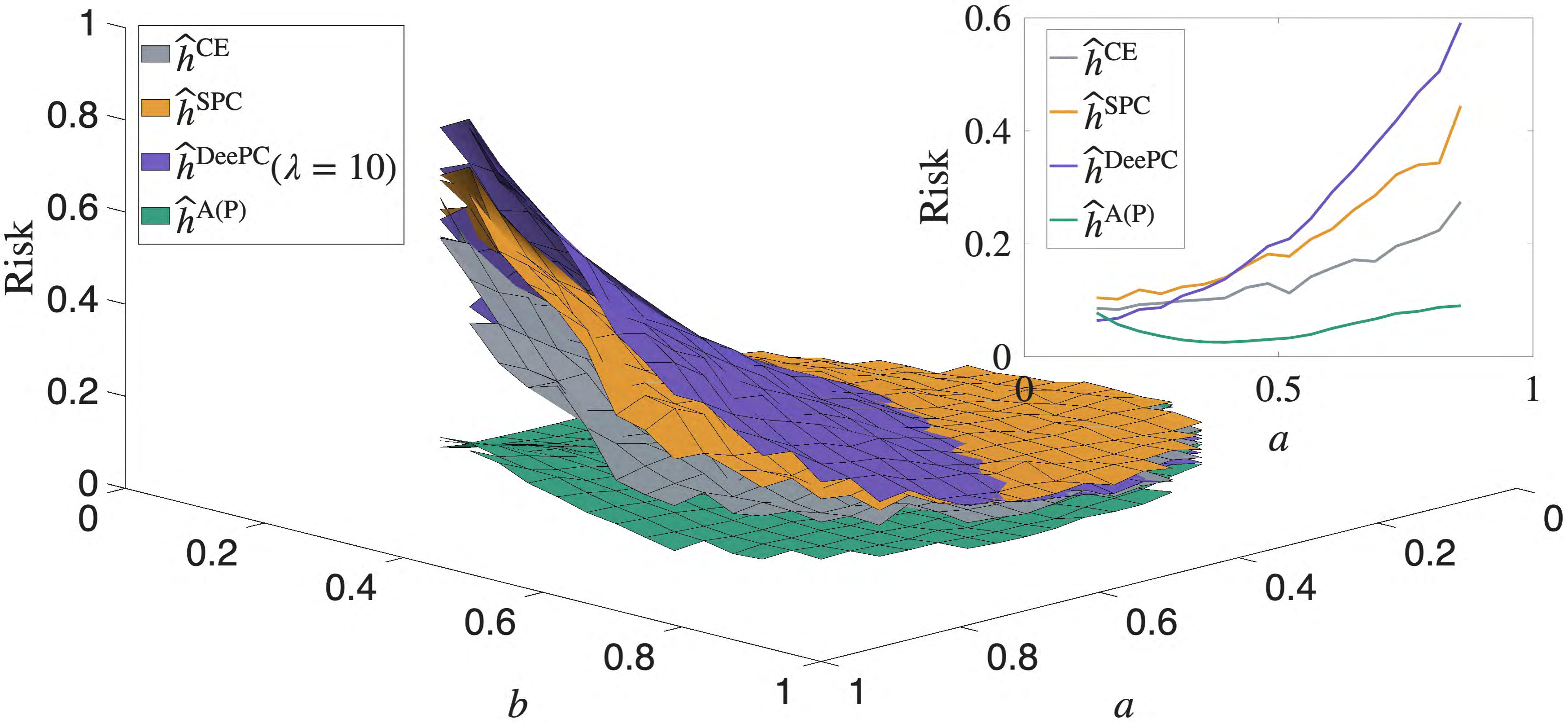}
	\caption{Pointwise risks $R_\theta(\hat h)$ of different controllers over $\mathcal D_\theta$. The slice corresponds to varying $a$ with $b=0.5$.}
	\label{fig:ex3-all-risks}
\end{figure}

\begin{figure}
	\centering
	\includegraphics[scale=0.12]{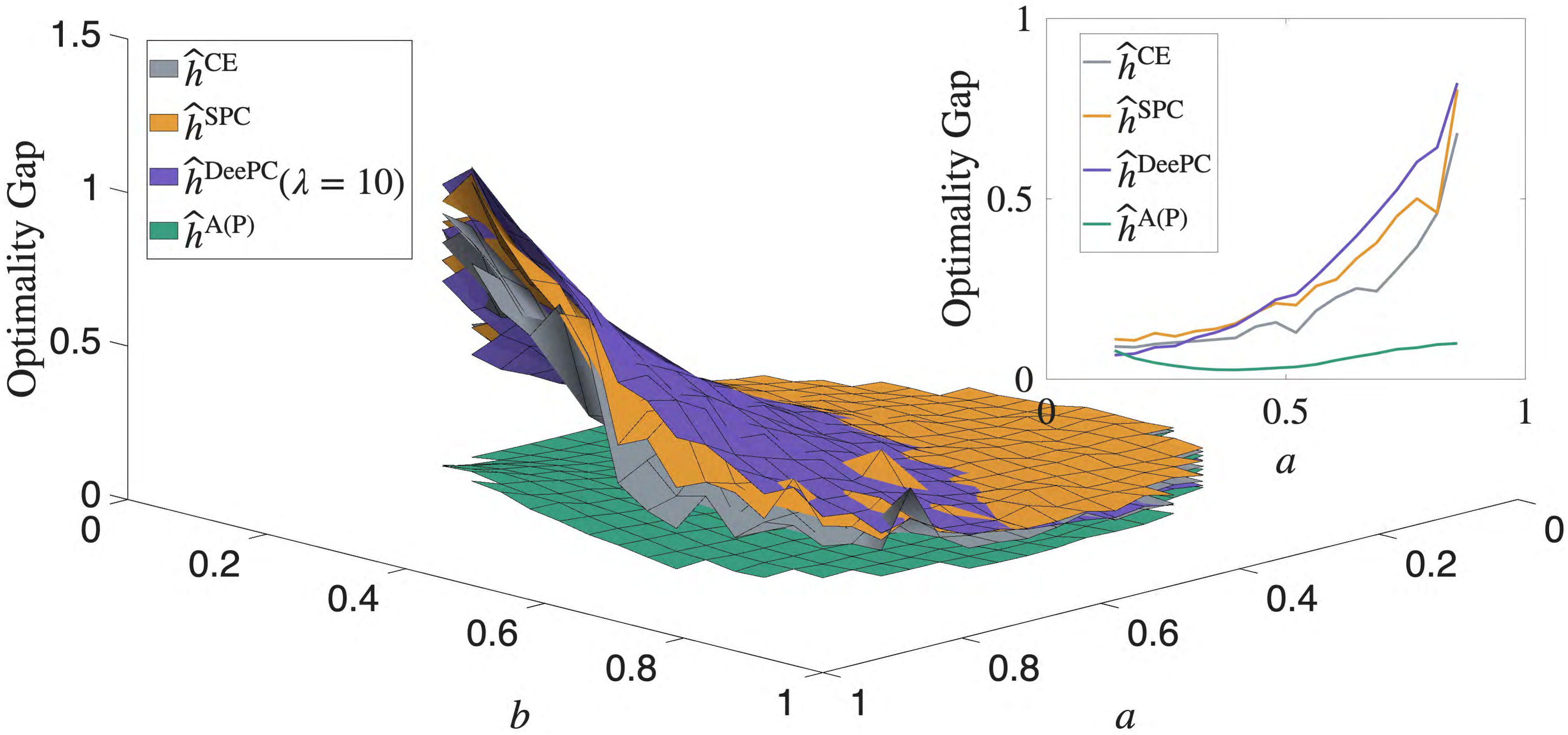}
	\caption{Suboptimality gaps of different controllers over $\mathcal D_\theta$. The slice corresponds to varying $a$ with $b=0.5$.}
	\label{fig:ex3-all-gaps}
\end{figure}

Figure~\ref{fig:ex3-all-risks} compares the pointwise risks of the four DDPC controllers introduced in Section~\ref{Sct3}. Unlike the bounds, the empirical risks fluctuate over $\mathcal D_\theta$, owing to the finite data length and the finite number of Monte Carlo trials; the overall trends are nevertheless consistent with the bounds. For instance, the risks of all controllers increase in the region $b\to0$, $a\to1$, in agreement with the bounds. Moreover, it can be observed that there is no uniformly best DDPC controller: $\hat h^{\mathrm{A(P)}}$ achieves the smallest risk $R_\theta (\hat{h})$ over most of the domain, reflecting its total risk minimization design, but $\hat h^{\mathrm{DeePC}}$ is better in subregions where $a\to0$. The average closed-loop suboptimality gaps in Figure~\ref{fig:ex3-all-gaps} follow the same pattern as the risks in Figure~\ref{fig:ex3-all-risks}, which supports the loss \eqref{eq:loss} as a surrogate for the realized suboptimality gap~\eqref{eq:cost_gap}.

\begin{figure}
	\centering
	\includegraphics[scale=0.12]{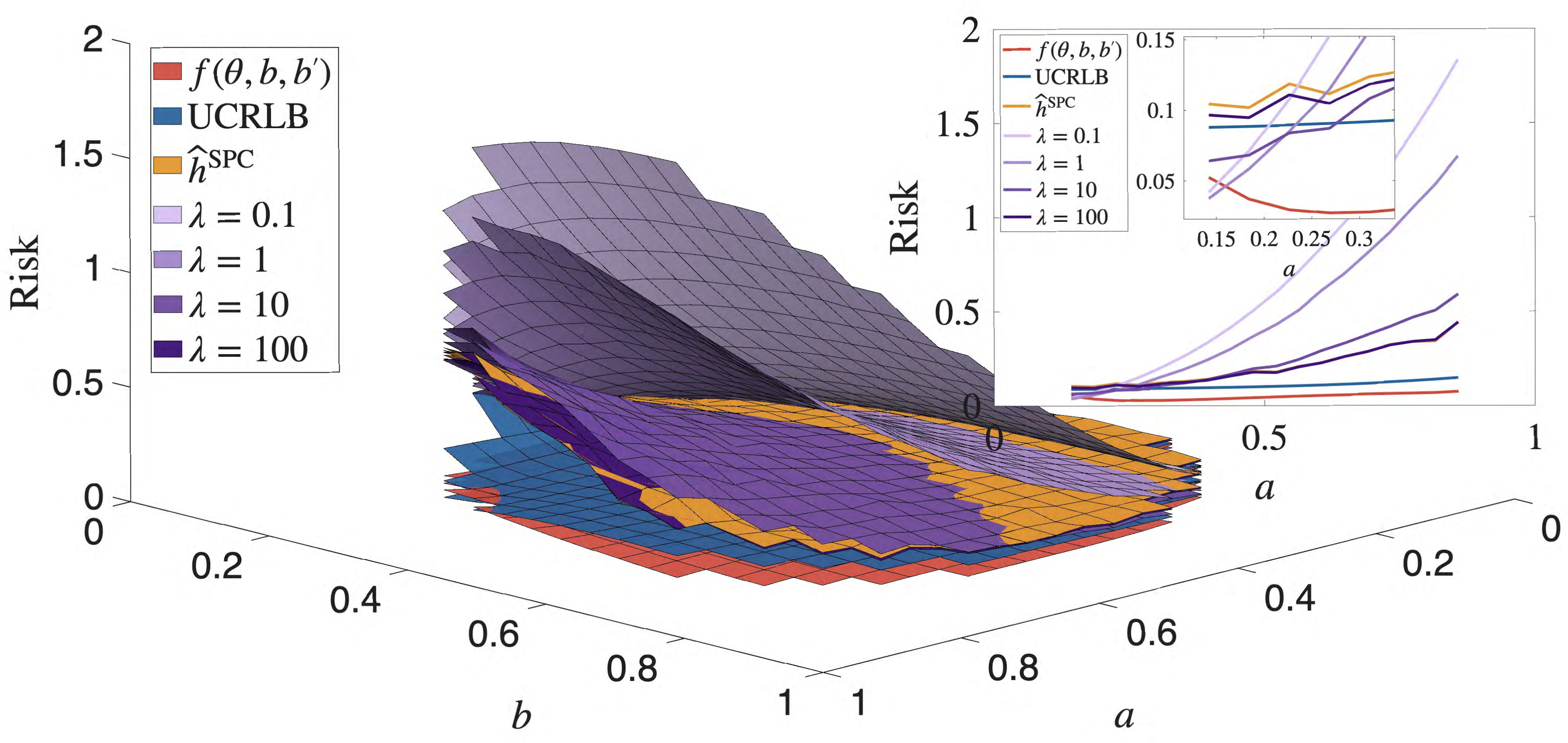}
	\caption{Pointwise risks of DeePC (purple) with $\lambda\in\{0.1,1,10,100\}$ and SPC (yellow), together with the UCRLB (blue) and the optimal-bias bound (red). The slices vary $a$ with $b=0.5$.}
	\label{fig:ex3-DeePC-reg}
\end{figure}

Figure~\ref{fig:ex3-DeePC-reg} illustrates the effect of regularization on the risk of DeePC. For every choice of $\lambda$, the total risk is subject to the same optimal-bias lower bound (red surface). As $\lambda$ increases, the risk of DeePC approaches that of SPC, consistent with \cite{Dorfler2022bridging}. The comparison also reveals a tradeoff: smaller values of $\lambda$ yield lower risk in regions where $a$ is small, but higher risk where $a$ is large. Thus, tuning $\lambda$ redistributes risk over the parameter domain, a point we revisit later.

%
%
%
%
%
%
%

\begin{figure}[!t]
	\centering
	\begin{subfigure}[t]{0.48\columnwidth}
			\centering
			\includegraphics[width=\linewidth]{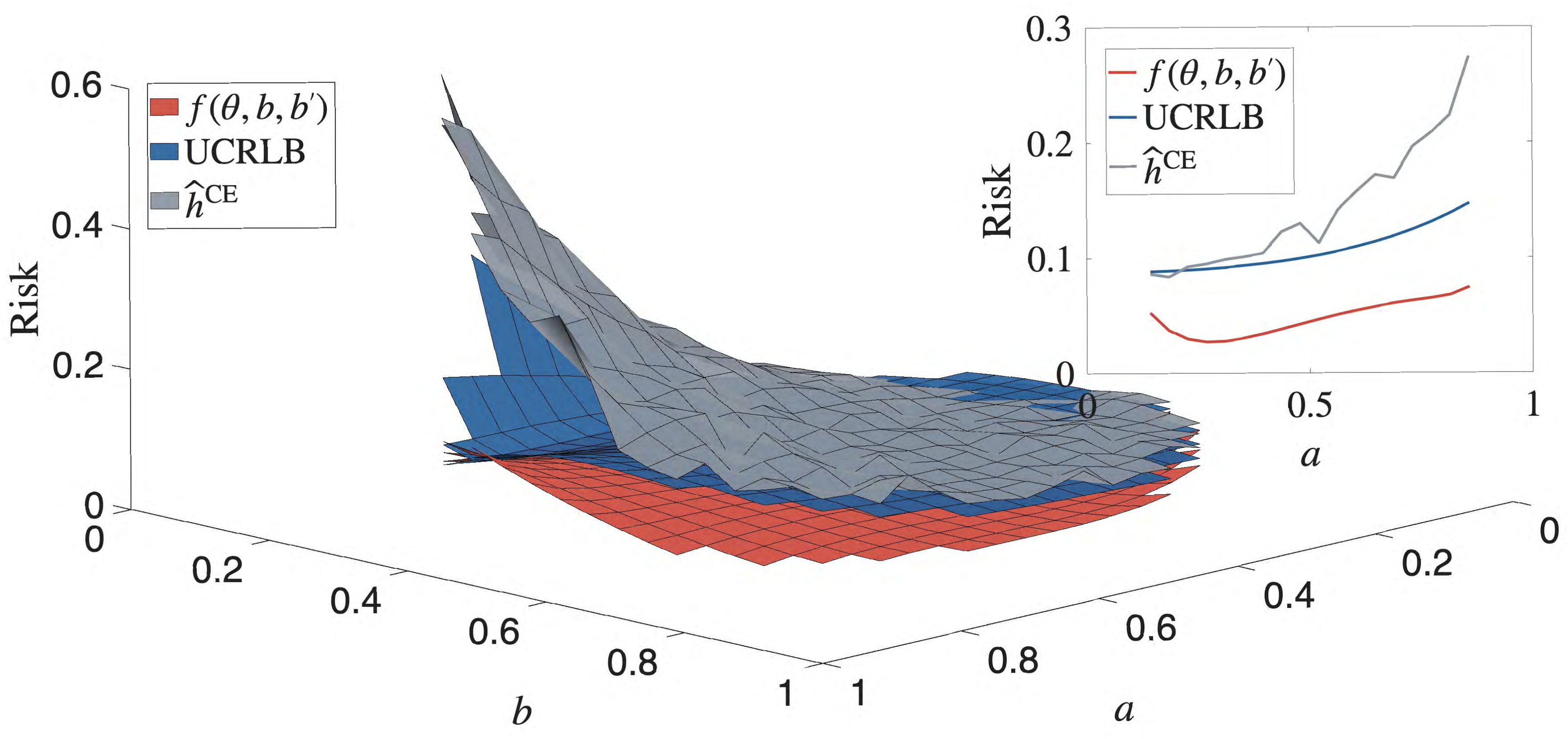}
			\caption{Pointwise risk $R_\theta(\hat h)$ of CE MPC}
			\label{fig:cep}
		\end{subfigure}
	\hfill
	\begin{subfigure}[t]{0.48\columnwidth}
			\centering
			\includegraphics[width=\linewidth]{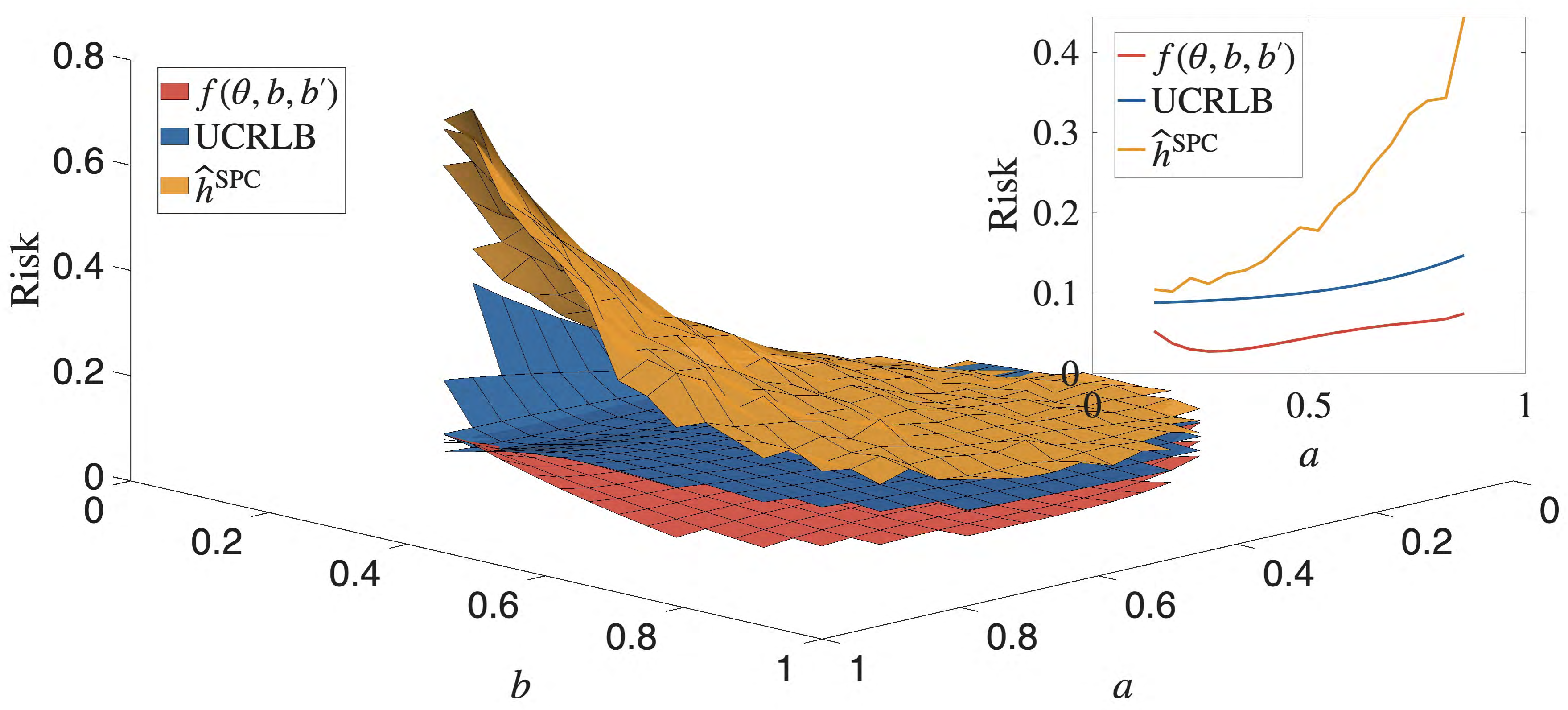}
			\caption{Pointwise risk $R_\theta(\hat h)$ of SPC}
			\label{fig:spc}
		\end{subfigure}
	
	\vspace{0.3em}
	
	\begin{subfigure}[t]{0.48\columnwidth}
			\centering
			\includegraphics[width=\linewidth]{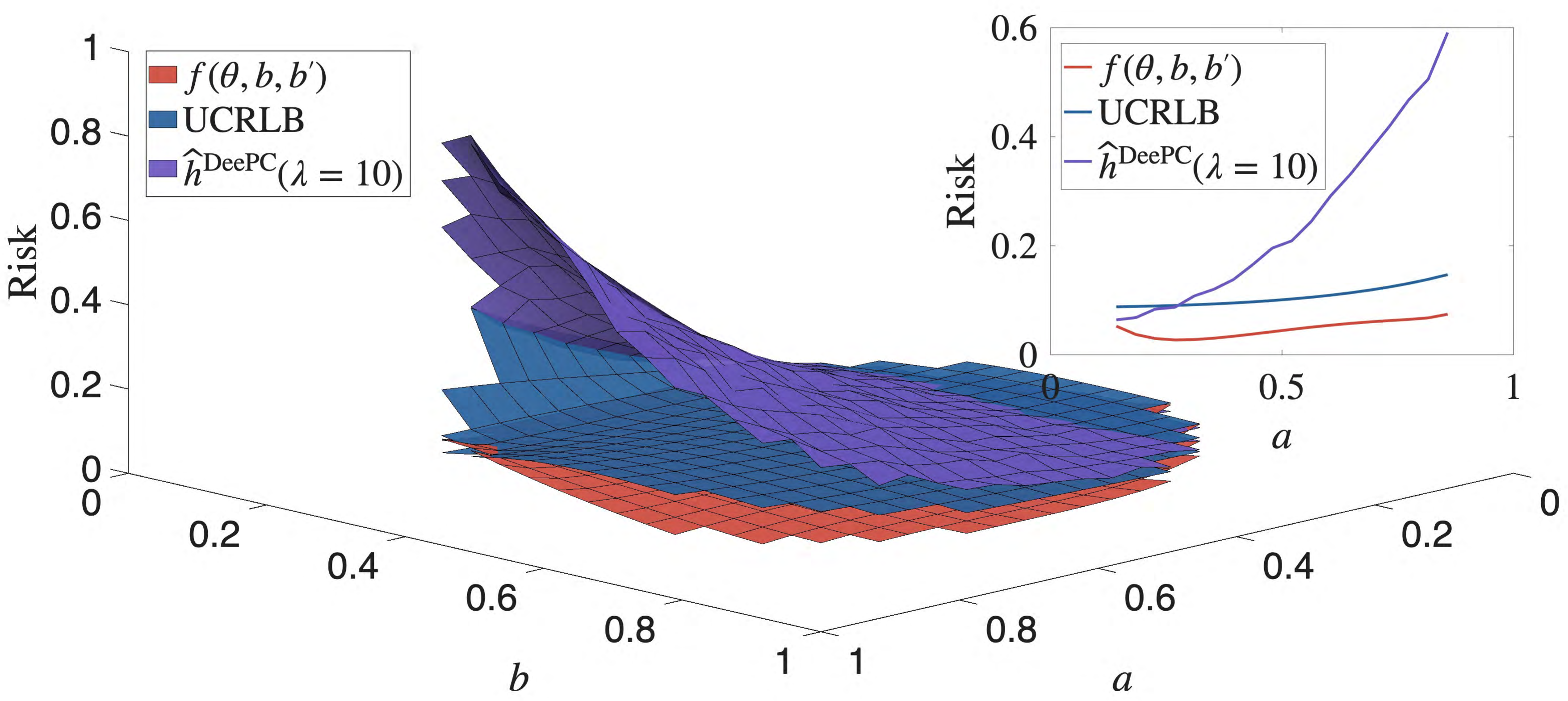}
			\caption{Pointwise risk $R_\theta(\hat h)$ of DeePC with $\lambda=10$}
			\label{fig:deepc}
		\end{subfigure}
	\hfill
	\begin{subfigure}[t]{0.48\columnwidth}
			\centering
			\includegraphics[width=\linewidth]{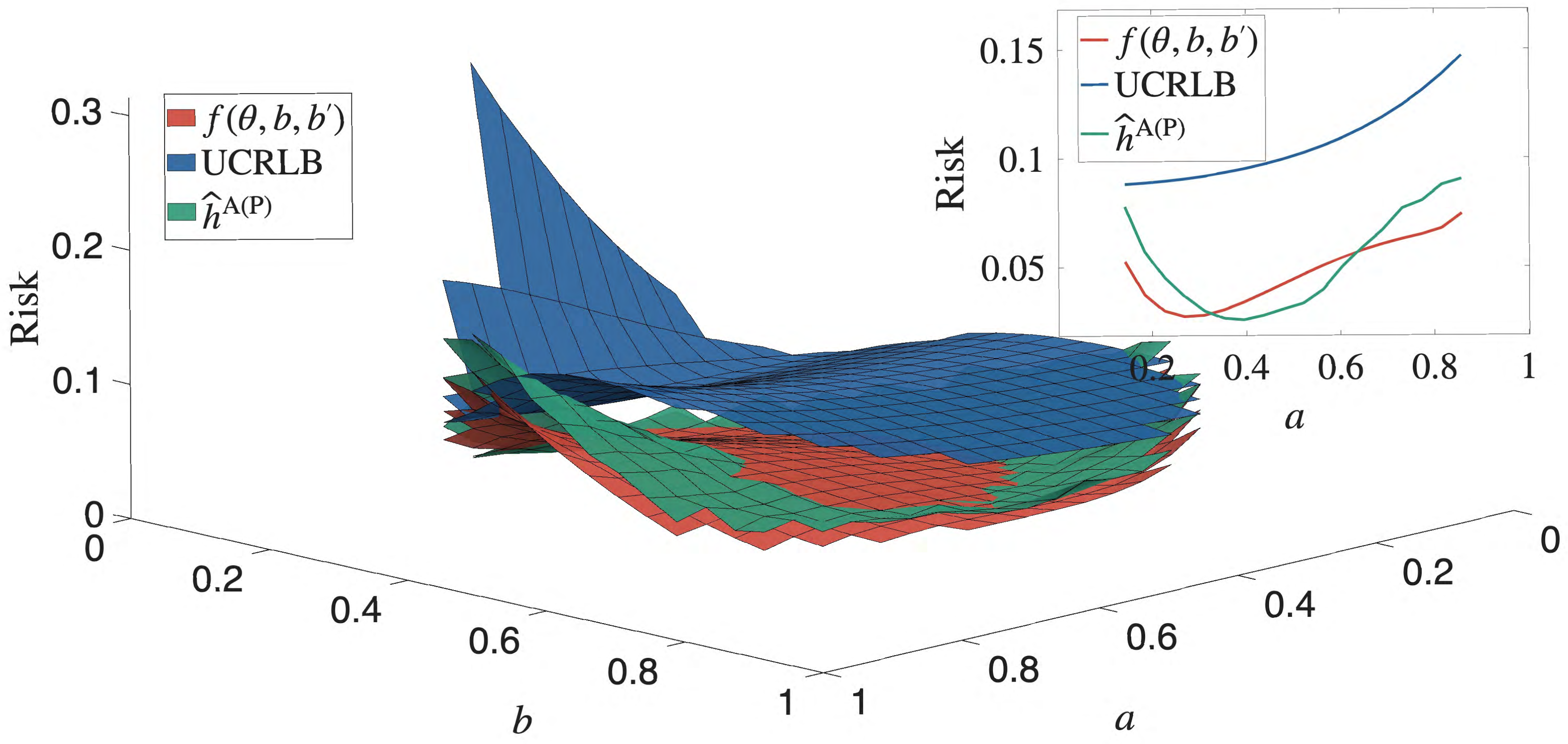}
			\caption{Pointwise risk $R_\theta(\hat h)$ of the total-risk minimization method}
			\label{fig:ap}
		\end{subfigure}
	
		\caption{Comparison of DDPC controllers v.s. the lower bounds over $\mathcal D_\theta$, with slices at $b=0.5$ along the direction $a\in[0.1,0.9]$.}
	\label{fig:LQR_comparisons}
\end{figure}

Figure~\ref{fig:LQR_comparisons} compares the pointwise risks of each controller with the bounds. The CEP controller (Figure~\ref{fig:cep}) follows the profile of the UCRLB and stays above it over much of the domain. Since the OLS estimate \eqref{eq:ls} is consistent, under some mild conditions, the resulting CEP controller $\hat h^{\text{CE}}$ is also consistent \cite[Proposition~1]{Ye2024online}. The UCRLB is therefore a relevant benchmark for $\hat h^{\text{CE}}$ when $N$ is large. SPC (Figure~\ref{fig:spc}) estimates the multi-step predictor consistently, but without imposing its structure: the redundant coefficients add variance, which appears as a larger risk over the whole domain. The same holds for DeePC with the projection-based regularizer (Figure~\ref{fig:deepc}), which is also consistent for the predictor when $\lambda$ is large. In addition, the regularization $\lambda = 10$ introduces a bias for finite $\lambda$, so the UCRLB is no longer informative for it and its risk should be read against $f(\theta,b,b')$. Moreover, the average-risk controller $\hat h^{\text{A(P)}}$ (Figure~\ref{fig:ap}) is biased by design: its risk follows $f(\theta,b,b')$ closely over the whole domain, and its total risk is the smallest of the four.

Finally, Figures~\ref{fig:deepc} and~\ref{fig:ap} illustrate the ``waterbed'' effect in DDPC. The controller $\hat h^{\text{DeePC}}$ achieves low risk for small $a$, while $\hat h^{\text{A(P)}}$ performs well near the center of the domain. Consistent with Figure~\ref{fig:ex3-DeePC-reg}, decreasing $\lambda$ further reduces the risk of DeePC for small $a$, at the expense of higher risk for larger $a$. These observations are analogous to the fundamental tradeoff in classical sensitivity shaping: Bode's sensitivity integral shows that reductions in sensitivity over certain frequency ranges must be compensated by increased sensitivity elsewhere \citep{Doyle2013feedback}. In the context of DDPC, reducing the risk w.r.t. the lower bound in one region of the parameter domain is accompanied by an increase in another, and the optimal-bias bound makes this redistribution explicit and quantifies how severe it is.

\vspace{-2mm}

\section{Conclusion} \label{Sct5}

In this work, we applied the statistical decision framework developed in \cite{He2026fundamental} to study fundamental limits of DDPC. We derived lower bounds on the pointwise and total risks and compared CE MPC, SPC, DeePC, and the total-risk minimization method through their explicit gain representations. The comparisons highlight the role of regularization in redistributing risk over the parameter domain. For DeePC, smaller regularization parameters improve performance for small $a$ at the expense of higher risk for larger $a$, motivating tuning according to the total risk over the domain of interest. Under the prescribed quadratic loss and parameter domain, the total-risk minimization method achieves the smallest total risk among all admissible decision rules. It therefore provides an optimal reference for assessing existing designs. The total-risk lower bound further quantifies unavoidable performance limitations and the associated ``waterbed'' effect. Future work will extend this method to online data-driven control.

\vspace{-2mm}
\bibliography{ifacconf}

\end{document}